\documentclass[preprint,aps,prc,nofootinbib,showpacs]{revtex4}

\usepackage{epsfig}
\usepackage{graphicx}

\begin{document}
\title{Model-independent methods to measure the $P$-parity of the $\Theta^+$-pentaquark in photoproduction experiments}
\author{Michail P. Rekalo}
\affiliation{\it National Science Center KFTI, 310108 Kharkov, Ukraine}
\author{Egle Tomasi-Gustafsson}
\affiliation{\it DAPNIA/SPhN, CEA/Saclay, 91191 Gif-sur-Yvette Cedex, 
France}
\date{\today}


\begin{abstract}
We discuss model independent methods to measure the $P$-parity of the $\Theta^+$-pentaquark, in the simplest processes of photoproduction, $\gamma+N\to \overline{K}+\Theta^+$, using definite relations between T-even polarization observables, which depend on the $P$-parity of the $\Theta^+$ baryon (with respect to the $NK$ system). One method, which holds for any photon energy and any $K$ meson production angle, is based on the relation between the $\Sigma_B$ asymmetry (induced by a linearly polarized photon beam, with unpolarized target) and the $D_{nn}$ component of the depolarization tensor (for unpolarized photon beam). Another method, which applies in collinear kinematics (or in threshold conditions), is related to the sign of the $D_{nn}$ component, with linearly polarized photon beam.

\end{abstract}
\maketitle
The experimental determination of the $P$-parity of the $\Theta^+$-pentaquark is important for the understanding of the underlying structure of this resonance  \cite{Di97,Ja03,St03,Ho03,Cs03,Sa03,Zh03a,Bi03}, therefore, several  suggestions concerning different processes, have been recently proposed in the  literature \cite{Zh03}.

The simplest photoproduction reaction of the $\Theta^+$-baryon, 
$\gamma+N\to \overline{K}+\Theta^+$, is in principle interesting for the determination of the $\Theta^+$ parity: all the observables (the differential cross section and the polarization effects) depend of course on the $P$-parity. However different models give different predictions on the angular and $E_{\gamma}$ behavior of the observables. Presently, the choice of a reliable model for the considered reaction is extremely difficult - due to the absence of the necessary experimental information: too much freedom is left in building a model, from the choice of an adequate approach to the knowledge of necessary coupling constants and phenomenological form factors. In this way, the determination of the P-parity would be ambiguous, because model dependent. 

A similar example is given by photo- end electro-production of strange particles, where a wide number of different models give a comparable description of the available data. Different reaction mechanisms with many unknown parameters and constants (to be determined from the data), have been suggested, but their predictive power is quite poor, and any new experimental information require, as a rule, to update the fitting procedure \cite{Sa03b}.

One can note that, in the first studies on photo- end electro-production of strange particles, both possibilities on the K-meson $P$-parity, $P(K)=\pm 1$, were systematically considered in the theoretical analysis \cite{Fa61}. Later on, many data on $\gamma+N\to K+Y$ and $e^-+N\to e^-+K+Y$,
($Y=\Lambda$ or $\Sigma$ hyperon) were collected. Their phenomenological interpretation relies on the negative $P$-parity of the $K$ meson (or, more exactly, the relative $P$-parity with respect to the $NY$ system). There is a so large arbitrariness in the theoretical approaches, that it is, {\it a priori},  commonly assumed, for simplicity, that the $ K$ meson $P$ parity is negative. Following the quark model, the $K$ meson is considered to be a pseudoscalar particle. But, in our opinion, the existing data and models on photo- and electro-production of strange particles can not be considered a proof of the pseudoscalar nature of the $K$-meson. The situation for $\gamma+N\to \overline{K}+\Theta^+$ is, in this respect, even more arbitrary, due to the lack of experimental data.

The same situation exists with charmed particles, mesons and baryons, because the $P$-parity of the lightest state, which is important for the identification of the parity of the excited charmed particles, is experimentally unknown. In the PDG tables \cite{PDG}, one can read: {\it "The parity of the lowest $\Lambda_c^+$ is defined to be positive, (as are the parities of the proton, neutron and $\Lambda$)"} - with respect to the $\Lambda_c$-hyperon - or {\it " $I,$ $J,$ $P$ need confirmation"} - concerning the charmed meson.

The problem of the $P$-parity of particles has to be solved independently on any model: parity is a fundamental property of a particle and affects the predictions of the models for particle photoproduction.

The basic question is therefore if it is possible, in general,  i.e. in a model independent way, to determine the $P$-parity of the particles produced in the reaction $\gamma+N\to M+B$, where $M$ is a meson with zero spin and $B$ is a baryon with spin 1/2. More exactly, we are considering the following processes:
\begin{eqnarray}
\gamma +N &\to &\overline{K}+\Theta^+,\nonumber \\
\gamma +N &\to &K+Y,~Y=\Lambda \mbox{~ or~} \Sigma\label{eq:reac},  \\
\gamma +N &\to &\overline{D}+Y_c,~Y_c=\Lambda_c \mbox{~ or~}\Sigma_c.\nonumber 
\end{eqnarray}
The purpose of this paper is to give model independent relations between polarization observables for all processes (\ref{eq:reac}), which allow, in principle,  to determine the quoted P-parities.

The idea to determine the $P$-parity of particles through relations between polarization observables is not new. Many years ago it has been suggested   \cite{Bi58,Bo59} to compare the analyzing power ${\cal A}$ in the processes:
\begin{eqnarray}
\pi^-+\vec p &\to &K^0 +\Lambda,\nonumber \\
K^-+\vec p &\to &\pi^0+\Lambda, \label{eq:reac1}   
\end{eqnarray}
induced by the proton target polarization, with the transversal polarization
${\cal P}$ of the produced $\Lambda$ hyperon (on unpolarized target), because the relation 
\begin{equation}
{\cal P}=-{\cal A} P(N\Lambda K)
\label{eq:mat}
\end{equation}
holds for any reaction mechanism, where $P(N\Lambda K)$ is the $K$ meson $P$-parity with respect to the $ N\Lambda$ reference frame. Let us recall here that one can not define an 'absolute' kaon $P$-parity, as in case of neutral particles,  $\gamma$, $\pi^0$, $\eta$, $\rho^0$, ..., because the kaon has nonzero hypercharge, and this quantum number conserves in strong and electromagnetic interactions. Such experiment, fundamental and relatively simple, has not yet been realized. 

More recently, other possibilities of model independent measurements of the kaon $P$-parity,  related to polarization phenomena in the processes $p+p\to K+p+\Lambda(\Sigma)$, in the near threshold region have been suggested  \cite{Pa99}.

These methods can also be applied to the determination of the $P$-parity of the $\Theta^+$ hyperon, through the study of polarization phenomena in the corresponding reactions of  $\Theta^+$-production in $pp$ collisions:
\begin{eqnarray}
p+ p &\to &\pi^+ + \Lambda^0 +\Theta^+,\nonumber \\
p+ p &\to &\overline{K}^0+p + \Theta^+. \label{eq:reac2}   
\end{eqnarray}
Again, in both reactions (\ref{eq:reac2}), we refer to the relative $\Theta^+$ parity, i.e. the $P$-parity of the $\pi\Lambda\theta^+$ or 
$p\overline{K}^0\Theta^+$ system, due to the conservation of strangeness in the production and the decay of $\Theta^+$.

Similarly, it is possible to determine the $\Theta^+$ $P$-parity in the photoproduction processes $\gamma +N \to \overline{K}+\Theta^+$ without any assumption  about the reaction mechanism, in different ways.

Due to the conservation of strangeness in the considered photoproduction process, the notion of absolute $\Theta^+$ $P$-parity has no physical meaning. Instead, one considers the relative parity - with respect to the N$\overline{K}$-system, $\Pi(\Theta)=P(\Theta N\overline{K})$. This definition has also the advantage to be free from any assumption about the $K$-meson parity. Moreover, all the dynamics of the process $\gamma +N \to \overline{K}+\Theta^+$ depends on the $\Theta^+$-parity, only through $\Pi(\Theta)$. Therefore, throughout the paper, the $\Theta^+$ $P$-parity, actually means $\Pi(\Theta)=P(\Theta N\overline{K})$, i.e. the relative parity of $\Theta^+$ with respect to the N$\overline{K}$-system. Such convention is coherent with the fact that $\Theta^+\to N\overline{K}$ is the main decay of the $\Theta^+$-hyperon.

Let us firstly consider the case of collinear kinematics for $\gamma +N \to \overline{K}+\Theta^+$, where helicity conservation results in one spin structure for the matrix element. The expression of the amplitude depends on the discussed $P$-parity \cite{Re03} and can be written as:
\begin{eqnarray}
{\cal M}_{col}^{(\pm)} &= &\chi_2^{\dagger}
{\cal F}_{col}^{(\pm)}\chi_1,\label{eq:eq4}\\
{\cal F}_{col}^{(+)} &= &\vec\sigma\cdot\vec\epsilon\times\hat{\vec k} 
f^{(+)}(E_{\gamma}),  \mbox{~if~} \Pi(\Theta) =+1,\label{eq:eq5}\\
 {\cal F}_{col}^{(-)} &= &\vec\sigma\cdot\vec\epsilon  f^{(-)}(E_{\gamma}),  \mbox{~if~} \Pi(\Theta) =-1,\label{eq:eq6}
\end{eqnarray}
where $\vec \epsilon$ is the real photon polarization vector, $\chi_1$ and $\chi_2$  are the
two-component spinors of the initial nucleon and the final baryon, 
$\hat{\vec k}$ is the unit vector along the three momenta of the photon beam (and of the $K-$meson) in the reaction center of mass system (CMS), $f^{(\pm)}(E_{\gamma})$ are the collinear amplitudes, where the upper indexes correspond to $P(N\Theta D)=\pm 1$, $\vec\epsilon \cdot\hat{\vec k} =0$.

Due to the presence of a single allowed amplitude in Eq. (\ref{eq:eq4}), all polarization phenomena have definite numerical values, which are independent on the model chosen for $f^{(\pm)}(E_{\gamma})$. Moreover all nonzero polarization effects take their maximal (absolute) value.

The spin structures Eqs. (\ref{eq:eq5}) and (\ref{eq:eq6}) are different, but as the single and double spin observables coincide, they can be distinguished only at the level of triple spin polarization correlations. This require the   measurement of  the dependence of the final baryon polarization on the polarization of the nucleon target, when the photon beam is linearly polarized.

Using Eqs. (\ref{eq:eq4}), (\ref{eq:eq5}) and (\ref{eq:eq6}), one can find:
\begin{equation}
{\cal P}_{2x}=-\Pi(\Theta){\cal P}_{1x},~{\cal P}_{2y}=-\Pi(\Theta) {\cal P}_{2x}, 
\label{eq:pxx}
\end{equation}
taking the $z$-axis along $\hat{\vec k}$ and the $x$-axis along the vector $\vec\epsilon$ of the photon linear polarization, when $\vec{\cal P}_{1,2}$ are the polarization vectors of the nucleon target and the produced $\Theta^+$.

Note that the same spin structure, Eqs. (\ref{eq:eq5}) and (\ref{eq:eq6}) describes the threshold amplitude, when the final particles are produced in S-state. In this case, the angular distribution of the emitted particles is isotropic in the CMS, and again, only one physical direction is defined. 

the second method we are suggesting holds for any kinematical conditions, i.e. for any $E_{\gamma}$ and $\cos\theta$ ($\theta$  is the $\overline{K}$ meson production angle). The following model independent relation holds between polarization observables in any reaction $\gamma+N\to M+B(1/2^{\pm})$ \cite{ETG03}:
\begin{eqnarray}
\Sigma_B(E_{\gamma},\cos\theta)&= & -P(BNM) D_{nn}(E_{\gamma},\cos\theta)\noindent\\
\Sigma_B(E_{\gamma},\cos\theta)&= &\displaystyle\frac{
\displaystyle\frac{d\sigma_{\perp}}{d\Omega}-
\displaystyle\frac{d\sigma_{\parallel}}{d\Omega}}
{
\displaystyle\frac{d\sigma_{\perp}}{d\Omega}+
\displaystyle\frac{d\sigma_{\parallel}}{d\Omega}
}
\label{eq:eq8}
\end{eqnarray}
where $P(BNM)$ is the relative parity of $B$, with respect to the $NM$-system, $d\sigma_{\perp}/{d\Omega}(d\sigma_{\parallel}/{d\Omega})$ is the differential cross section for $\vec\gamma+N\to M+B$ with a linearly polarized photon beam, when the three vector $\vec\epsilon$ is transversal (parallel) to the reaction plane, and $D_{nn}$ is a specific component of the depolarization tensor, which characterizes the $B$-polarization normal to the reaction plane, in collisions of unpolarized photon beam with a nucleon target polarized normally to the reaction plane.

Note that in collinear kinematics, or at the reaction threshold, the relation 
(\ref{eq:eq8}) is equivalent to the identity $0=0$, because, in such conditions the reaction plane can not be defined, so that both polarization observables, $\Sigma_B$ and $D_{nn}$ are proportional to $\sin^2\theta$ ($\to 0$), independently on the reaction mechanism.

One can see that the model independent methods (which allow to determine the $\Theta^+$ $P$-parity in $\gamma +N \to \overline{K}+\Theta^+$), Eqs. (\ref{eq:pxx}) and Eq. (\ref{eq:eq8}), require the measurement of the polarization of the produced baryon. In case of associative production of strange ($K+Y$) or charmed ($\overline{D}+\Lambda_c$) particles, such measurements are relatively easy, because the hyperons $Y$ ($\Lambda$ or $\Sigma$) and $\Lambda_c^+$, decaying through weak interaction, are self analyzing particles. Therefore for charm photoproduction such measurements can be done by the running COMPASS experiment \cite{COMPASS}, with a polarized target and a muon beam, which can generate linearly polarized photons. 

In case of $ \Theta^+$ photoproduction, the situation is more difficult, because the polarization of the $ \Theta^+$ baryon can be measured only through the measurement of the proton polarization in the decay $\Theta^+\to p+K^0$. Therefore we agree with Ref. \cite{Thomas}: "{\it  Even under these ideal conditions, the decay angular distribution of this strongly decaying particle gives information only on the spin and not on the parity, unless the polarization of the final nucleon is measured}", but, contrary to  Ref. \cite{Thomas}, we showed here that the polarization of $ \Theta^+$ in photoproduction reactions, may give access to the $ \Theta^+$ parity, in a model independent way. We suggested above two different methods, one of which requires the measurement of the final nucleon polarization, to determine the $ \Theta^+$ parity in $\gamma +N \to \overline{K}+\Theta^+$. These methods represent the theoretical solution of the problem of the model-independent determination of the $ \Theta^+$ parity. The realization of the suggested experiments may be difficult from the experimental point of view, but in the literature one can find only model-dependent methods, not necessarily more simple.


Let us mention that the model independent methods, based on Eq. \ref{eq:mat}, can be applied also for the measurement of the $P$-parity of $\Theta^+$ and charm particles, through other binary reactions:
\begin{equation}
K+N\to \pi+\Theta^+,~\pi+N\to \overline{K}+\Theta^+,~\pi+N\to\overline{D}+Y_c.
\label{eq:eq9}
\end{equation}
But, again, it is a difficult problem for the $\Theta^+$ baryon, because its transversal polarization has to be compared with the analyzing power. 

For charmed particles, this method can be considered the most simple, as it involves the measurement of single-spin observables only for the determination of two T-odd polarization observables. The two independent amplitudes, which characterize the general spin structure of the matrix elements for each of the processes (\ref{eq:eq9}), have to be complex, with nonzero relative phase.  Such condition is {\it a priori} satisfied for $\Theta^+$ production in $\pi N$ and $KN$-collisions, in the near threshold region, where the S-channel contributions of different $N^*$ or $Y^*$ resonances can generate large T-odd polarization effects. Evidently, for charm particle production, such resonance mechanisms are not working, but the necessary phase could be generated by $D$ and $D^*$ Regge contributions, for example.

In conclusion, the model independent determination of the $\Theta^+$ parity could be realized, in principle, using definite relations between polarization observables, in reactions as $ \gamma+N\to \overline{K}+\Theta^+$,
$K+N\to\pi+\Theta^+$, $\pi+N\to\overline{K}+\Theta^+$. These procedures are based on the same physics as the corresponding methods suggested for strange or charm particle production, but in the first case, they involve difficult experiments, which require the measurement of the $\Theta^+$ polarization. 

Another suggestion, which allows to avoid this problem, can be done, using again the analogy with strange particle production. The reactions $p+p\to \pi^+ +\Lambda^0+\Theta^+$, $p+p\to \overline{K}^0+p+\Theta^+$, which are similar to $p+p\to K^++\Lambda+p$ \cite{Pa99}, can be used for the determination of the $P$-parity of the $\Theta^+$ baryon. The same polarization phenomena can be considered for the process 
$p+p\to\Theta^+ +\Sigma^+$, where instead of the $\Theta^+$ polarization, the polarization of the $\Sigma^+$ hyperon (a self analyzing particle) is more experimentally accessible.

These methods are, in general, equivalent, but in case of $pp$-collisions there are different possibilities to avoid the measurement of the $\Theta^+$ polarization, considering the collisions of both polarized protons or the polarization transfer coefficients, from a polarized proton beam (with unpolarized proton target) to the produced $\Lambda$ or $\Sigma$ hyperons.
This will be discussed in detail in a forthcoming paper.

We are grateful to A. Gal for interesting remarks on methods to measure the kaon parity, and to K. Kakayama for driving our attention to a precise definition of the parity, for the $\Theta^+$-hyperon.

{}

\end{document}